# 6G White Paper on Edge Intelligence

## Abstract

In this white paper we provide a vision for 6G Edge Intelligence. Moving towards 5G and beyond the future 6G networks, intelligent solutions utilizing data-driven machine learning and artificial intelligence become crucial for several real-world applications including but not limited to, more efficient manufacturing, novel personal smart device environments and experiences, urban computing and autonomous traffic settings. We present edge computing along with other 6G enablers as a key component to establish the future 2030 intelligent Internet technologies as shown in this series of 6G White Papers.

In this white paper, we focus in the domains of edge computing infrastructure and platforms, data and edge network management, software development for edge, and real-time and distributed training of ML/AI algorithms, along with security, privacy, pricing, and end-user aspects. We discuss the key enablers and challenges and identify the key research questions for the development of the Intelligent Edge services. As a main outcome of this white paper, we envision a transition from Internet of Things to *Intelligent Internet of Intelligent Things* and provide a roadmap for development of *6G Intelligent Edge*.

## Authors

**Ella Peltonen**, University of Oulu, Finland, ella.peltonen@oulu.fi
**Mehdi Bennis**, University of Oulu, Finland, mehdi.bennis@oulu.fi
**Michele Capobianco**, Capobianco, Italy, michele@capobianco.net
**Merouane Debbah**, Huawei, France, merouane.debbah@huawei.com
**Aaron Ding**, TU Delft, Netherlands, aaron.ding@tudelft.nl
**Felipe Gil-Castiñeira**, University of Vigo, Spain, xil@gti.uvigo.es
**Marko Jurmu**, VTT Technical Research Centre of Finland, Finland, marko.jurmu@vtt.fi
**Teemu Karvonen**, University of Oulu, Finland, teemu.3.karvonen@oulu.fi
**Markus Kelanti**, University of Oulu, Finland, markus.kelanti@oulu.fi
**Adrian Kliks**, Poznan University of Technology, Poland, adrian.kliks@put.poznan.pl
**Teemu Leppänen**, University of Oulu, Finland, teemu.leppanen@oulu.fi
**Lauri Lovén**, University of Oulu, Finland, lauri.loven@oulu.fi
**Tommi Mikkonen**, University of Helsinki, Finland, tommi.mikkonen@helsinki.fi
**Ashwin Rao**, University of Helsinki, Finland, ashwin.rao@helsinki.fi
**Sumudu Samarakoon**, University of Oulu, Finland, sumudu.samarakoon@oulu.fi
**Kari Seppänen**, VTT Technical Research Centre of Finland, Finland, kari.seppanen@vtt.fi
**Paweł Sroka**, Poznan University of Technology, Poland, pawel.sroka@put.poznan.pl
**Sasu Tarkoma**, University of Helsinki, Finland, sasu.tarkoma@helsinki.fi
**Tingting Yang**, Pengcheng Laboratory, China, yangtt@pcl.ac.cn

## Acknowledgements

This draft white paper has been written by an international expert group, led by the Finnish 6G Flagship program (www.6gflagship.com) at the University of Oulu, within a series of twelve 6G white papers to be published in their final format in June 2020.

# 1. Introduction

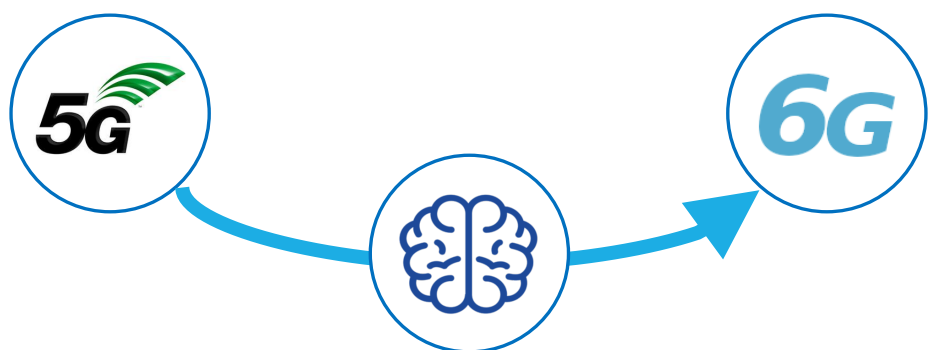

**FIGURE 1: THE TRANSITION FROM 5G TO 6G ENABLED BY EDGE INTELLIGENCE.**

**Edge Intelligence (EI), powered by Artificial Intelligence (AI) techniques (e.g. machine learning, deep neural networks etc.), is already being considered to be one of the key missing elements in 5G networks** and will most likely represent a key enabling factor for future 6G networks, to support their performance, their new functions, and their new services. Consequently, this whitepaper aims to provide an overarching understanding on why edge intelligence is an important aspect in 6G and what are the leading design principles and technological advancements that are guiding the work towards the edge intelligence for 6G.

In the last few years, we have witnessed a growing market and exploitation of AI solutions in a wide spectrum of ICT applications. AI services are becoming more and more popular in various ways, including intelligent personal assistants, video/audio surveillance, smart city operations, and autonomous vehicles. In fact, entire industries are taking new forms -- a prime example is Industry 4.0 that aims to digitize manufacturing, robotics, automation, and related industrial fields as a part of digital transformation. Furthermore, the increasing use of computers and software calls for new types of tradeoffs in designs, concerning for instance energy and timing constraints of computations and data transmissions as well as privacy and security.

The increased interest in AI can be attributed to recent phenomena, high-performance yet affordable computing and increasing amount of data, generated by various ubiquitous devices from personal smartphones to industrial robots. Powerful and low-cost processing and storage resources of cloud computing are available for anyone with a credit card, and there the abundance of resources meets the hungry requirements of AI, called to elaborate enormous quantities of big data. Furthermore, the **high density of base stations in megacities (and high density of devices) provide a good basis for edge and fog computing.**

The devices generating and consuming the data are commonly located at the edge of the networks, near the users and systems under monitoring, surveillance, or control. However, this megatrend has received only little attention. Indeed, wide diffusion of smart terminals, devices, and mobile computing, the Internet of Things (IoT) and the proliferation of sensors and video cameras are generating several hundreds of ZB data at the network edge. Furthermore, increasing use of machine learning models with small memory footprint - such as TinyML - that operate at the edge plays an important role. Taking this into account in computational models means that the centralized cloud computing model needs to be extended towards the edge.

**Edge Computing (EC) is a distinguished form of cloud computing that moves part of the service-specific processing and data storage from the central cloud to edge network nodes, physically and logically close to the data providers and end users.** Among the expected benefits of edge computing deployment in current 5G networks there are: performance improvements, traffic optimization, and new ultra-low-latency services. Edge intelligence in 6G will significantly contribute to all these above-mentioned aspects. Moreover, edge intelligence capability will enable development of a whole new category of products and services. New

business and innovation avenues around edge computing and edge intelligence are likely to emerge rapidly in several industry domains. Sometimes, the term Fog Computing is also used, to highlight that in addition to running things at the edge, also computers located between the edge device and the central cloud are used. While various definitions for edge and fog computing with subtle differences exist, we use the terms interchangeably to denote flexible executions that are run in computers outside the central cloud.

One definition of particular importance for 5G and beyond systems is given by the Multi-access Edge Computing (MEC) initiative within ETSI[1]. In this architecture, a mobile edge host runs a mobile edge platform that facilitates the execution of applications and services at the edge. The ETSI MEC standard connects the MEC applications and services with the cellular domain through the standardized APIs, such as access to base station information and network slicing support. From the data analytics point of view edge intelligence refers to data analysis and development of solutions at or near the site where the data is generated and further utilized. By doing so, edge intelligence allows reducing latency, costs, and security risks, thus making the associated business more efficient. From the network perspective, edge intelligence mainly refers to intelligent services and functions deployed at the edge of network, probably including the user domain, the tenant domain, or close to the user or tenant domain or across the boundary of network domains.

In its basic form edge intelligence involves an increasing level of data processing and capacity to filter information on the edge. However, intelligence is defined "a priori". With increasing levels of artificial intelligence at the edge, it is possible to bring some AI features to each node, as well as on clusters of nodes, so that they can learn progressively and possibly share what they learn with other similar (edge) nodes to provide, collectively, new added value services or optimized services. **Hence, it can be predicted that the evolution of telecom infrastructures towards 6G will consider highly distributed AI, moving the intelligence from the central cloud to edge computing resources.** Target systems include advanced IoT applications and digital transformation projects. Furthermore, edge intelligence is a necessity for a world where intelligent autonomous systems are commonplace, in particular when considering situations where machines and humans cooperate (such as working environments) due to safety reasons.

At the moment, software and hardware optimized for edge intelligence are in their infancy and we are seeing an influx of edge devices such as Coral[2] and Jetson[3] that are capable of performing AI computation. Regardless, current AI solutions are resource- and energy-hungry and time-consuming. In fact, many commonly used machine learning and deep neural network algorithms still rely on Boolean algebra transistors to do an enormous amount of digital computations over massive-scale data sets. In the future, the number and size of available data sets will only increase whilst AI performance requirements will be more and more stringent, for expected (almost) real-time ultra-low latency applications. We see this trend cannot be really sustainable in the long term.

To give a concrete example of non-optimal hardware and software in 5G, we remind that in basic functioning of especially Deep Neural Networks (DNN), each high-level layer learns increasingly abstract higher-level features, providing a useful, and at times reduced, representation of the features to a lower-level layer. A roadblock is that chipsets technologies are not becoming faster at the same pace as AI solutions are progressing in serving markets' expectations and needs. Nanophotonic technologies could help in this direction: DNNs operations are mostly matrix multiplication, and nanophotonic circuits can make such operations almost at the speed of light and very efficiently due to the nature of photons. In common language, photonic/optical computing uses electromagnetic signals (e.g., via laser beams) to store, transfer, and process information. Optics has been around for decades, but until now, it has been mostly limited to laser transmission over optical fiber. Nanophotonic technologies, using optical signals to do computations and store data, could accelerate AI computing by orders of magnitude in latency, throughput, and power efficiency.

---

[1] https://www.etsi.org/technologies/multi-access-edge-computing
[2] https://www.coral.ai/
[3] https://developer.nvidia.com/buy-jetson

In-memory computing is a promising approach to addressing the processor-memory data transfer bottleneck in computing systems. In-memory computing is motivated by the observation that the movement of data from bit-cells in the memory to the processor and back (across the bit-lines, memory interface, and system interconnect) is a major performance and energy bottleneck in computing systems. Efforts that have explored the closer integration of logic and memory are variedly referred to in the literature as logic-in-memory, computing-in-memory and processing-in-memory. These efforts may be classified into two categories – moving logic closer to memory, or near memory computing, and performing computations within memory structures, or in-memory computing [1]. In memory computing appears to be a suitable solution to support the hardware acceleration of DNN. System-on-Chip architectures like Adaptive Computing Acceleration Platform (ACAP) are yet another approach for AI applications. ACAPs integrate generic CPUs with AI and DSP specific engines as well as programmable logic in a single device. Internal memory and high-speed interconnection networks make it possible to implement the whole AI processing pipeline within a single device eliminating the need of transferring data off-the-chip [2].

Software supporting AI development is also one of the understudied aspects of the current 5G development. Tools, methods and practices we use to build edge devices, cloud software, gateways that connect them, and end-user applications are diverging due to various reasons, including performance, memory constraints, and productivity. This means that the responsibilities of different devices are still largely "a priori" defined during their design and implementation, and that we are far from software capabilities that would allow software to "flow" from one device to another (so-called liquid software). **Without liquid software as a part of the future 6G networks, we are stuck with an approach where we have to decide where to locate the intelligence at the network topology at design time due, since the computations cannot be easily relocated without design-time preparations.**

In this white paper, we aim to shed light to the challenges of edge AI, potential solutions to these challenges, and a roadmap towards intelligent edge AI. The paper is structured as follows. In Section 2, we discuss related work to motivate the paper. In Section 3, we provide an insight to our vision of edge AI. In Section 4, we address challenges and key enablers of edge AI in the context of the emerging era of 6G. In Sections 5 and, we present key research questions and a roadmap to meet the vision.

## 2. Related Work

Vision-oriented and positioning papers on 6G edge intelligence are starting to emerge. Zhou et al. [3] and Xu et al. [28] conduct a comprehensive survey of the recent research efforts on Edge Intelligence. Specifically, they review the background and motivation for artificial intelligence running at the network edge, concentrating on Deep Neural Networks (DNN), a popular architecture for supervised learning. Further, they provide an overview of the overarching architectures, frameworks, and emerging key technologies for deep learning model towards training and inference at the network edge. Finally, they discuss the open challenges and future research directions on edge intelligence.

Rausch and Dustdar [4] investigate the trends and the possible "convergence" between Humans, Things, and AI. In their article, they distinguish three categories of edge intelligence use cases: public such as smart public spaces, private such as personal health assistants and predictive maintenance (corporate) and intersecting such as autonomous vehicles. It is unclear who will own the future fabric for edge intelligence, whether utility-based offerings for edge computing will take over as is the case in cloud computing, whether telecommunications will keep up with the development of mobile edge computing, what role governments and the public will play, and how the answers to these questions will impact engineering practices and system architectures.

To address the challenges for data analysis of edge intelligence, computing power limitation, data sharing and collaborating, and the mismatch between the edge platform and AI algorithms, Zhang et al. [5] introduce an Open Framework for Edge Intelligence (OpenEI) which is a lightweight software platform to equip the edge with intelligent processing and data sharing capability. Similarly, the ARM compute library[4], the Qualcomm Neural Processing SDK[5], the Xilinx Vitis AI[6], and Tensorflow lite[7] offer solutions for performing AI computations on low power devices that can be deployed at the edge. More in general, the experience of Edge Computing is worth recalling. Mohan [6] adopts an Edge Computing Service Model based on a hardware layer, an infrastructure layer, and a platform layer to introduce a number of research questions. Hamm et al. [7] present an interesting summary based on the consideration of 75 Edge Computing initiatives. The Edge Computing Consortium Europe (ECCE)[8] aims at driving the adoption of the edge computing paradigm within the manufacturing and other industrial markets with the specification of a Reference Architecture Model for Edge Computing (ECCE RAMEC), the development of reference technology stacks (ECCE Edge Nodes), the identification of gaps and recommendation of best practices by evaluating approaches within multiple scenarios (ECCE Pathfinders).

On the theoretical side, Park et al. [8] highlights the need for distributed, low-latency and reliable machine learning at the wireless network edge to facilitate the growth of mission-critical applications and intelligence devices. Therein, the key building blocks of machine learning at the edge are laid out by analyzing different neural network architectural splits and their inherent tradeoffs. Furthermore, Park et al. [8] provides a comprehensive analysis of theoretical and technical enablers for edge intelligence from different mathematical disciplines and presents several case studies to demonstrate the effectiveness of edge intelligence towards 5G and beyond.

In a series of position papers, Lovén et al. [9] have divided Edge AI in Edge for AI, comprising the effect of the edge computing platform on AI methods, and AI for Edge, comprising how AI methods can help in the orchestration of an edge platform. They identify communication, control, security, privacy and application verticals as the key focus areas in studying the intersection of AI and edge computing, and outline the architecture for a secure, privacy aware platform which supports distributed learning, inference and decision making by edge-native AI agents.

---

[4] https://developer.arm.com/ip-products/processors/machine-learning/compute-library
[5] https://developer.qualcomm.com/software/qualcomm-neural-processing-sdk
[6] https://www.xilinx.com/products/design-tools/vitis/vitis-ai.html
[7] https://www.tensorflow.org/lite/
[8] https://ecconsortium.eu/

Almost identically to Lovén et al. [9], Deng et al. [10] separate AI for Edge and AI on Edge. In their study, Deng et al. discuss the core concepts and a research road-map to build the necessary foundations for future research programs in edge intelligence. AI for Edge is a research direction focusing on providing a better solution to the constrained optimization problems in edge computing with the help of effective AI technologies. Here, AI is used for enhancing edge with more intelligence and optimality, resulting in Intelligence-enabled Edge Computing (IEC). AI on Edge, on the other hand, studies how to carry out the entire lifecycle of AI models on edge. It is a paradigm of running AI model training and inference with device-edge-cloud synergy, with an aim on extracting insights from massive and distributed edge data with the satisfaction of algorithm performance, cost, privacy, reliability, efficiency, etc. Therefore, it can be interpreted as Artificial Intelligence on Edge (AIE).

# 3. Vision for the 2030s Edge-driven Artificial Intelligence

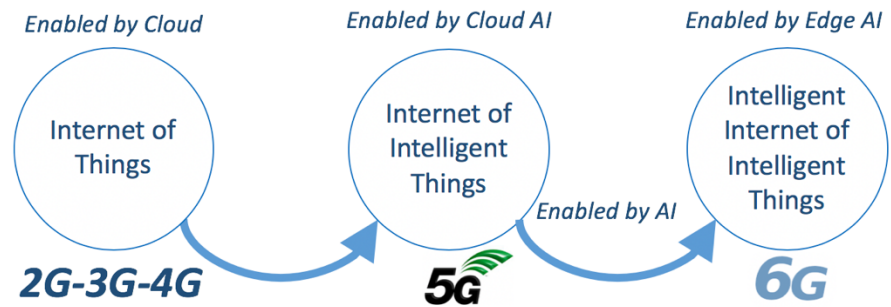

FIGURE 2: EVOLUTION OF THE "INTELLIGENT INTERNET OF INTELLIGENT THINGS"

**There is virtually no major industry where modern artificial intelligence is not already playing a role.** That is especially true in the past few years, as data collection and analysis has ramped up considerably thanks to robust IoT connectivity, the proliferation of connected devices and ever-speedier computer processing. Regardless the impact artificial intelligence is having on our present day lives, it is hard to ignore that in the future it will be enabling new and advanced services for: (i) Transportation and Mobility in three dimensions, (ii) Manufacturing and Industrial Maintenance, (iii) Healthcare and wellness, (iv) Education and Training, (v) Media and entertainment, (vi) eCommerce and Shopping, (vii) Environmental Protection, (viii) Customer Services. The complexity of the resulting functionalities requires an increasing level of distributed intelligence at all levels to guarantee efficient, safe, secure, robust and resilient services.

Similarly, to the transition we are experiencing from Cloud to Cloud Intelligence, we are constantly assisting at an evolution from the "Internet of Things" to the "Internet of Intelligent Things". Given the requirements above, what is more and more evident is also the need for an **"Intelligent Internet of Intelligent Things"** to make such internet more reliable, more efficient, more resilient, and more secure. **This is exactly the area where 6G communication with Edge-driven artificial intelligence can play a fundamental role.**

Compared with edge computing efforts from cloud service providers such as Google, Amazon, and Microsoft, there is a tighter integration advantage of computing and communication in 6G by telecom operators. For instance, 6G base stations can be a natural deployment of edge intelligence that requires both computing and communication resources. This is likely to represent a new opportunity for telecom operators and, to some extent, tower operators, to regain centrality in the market and to increase the added value of their offer.

When the connected objects become more intelligent in the 6G era, it is hard to believe that we can deal with them and with the complexity of their use and of their working conditions by continuing using the communication network in a static, simplistic, and dumb manner. The same need will likely emerge for any other services using the future communication networks, including phone calls, video calls, video conferences, video on demand, augmented and mixed reality video streaming, where the wireless communication network will no longer just provide a "connection" between two or more people or a "video channel" on demand from a remote repository to the user's TV set, but will bring the need to properly authenticate all involved parties, guarantee the security of data fluxes eventually using a dedicated blockchain, and recognizing in real time unusual or abnormal behavior. Data exchange will in practice be much more than just pure data exchange, but will exchange a number of past, present, and possibly future properties of those data. In the future 6G wireless communication networks, trust, service level, condition monitoring, fault detection, reliability, and resilience will define fundamental requirements, and artificial intelligence solutions are extremely promising candidates to play a fundamental role to satisfy such requirements.

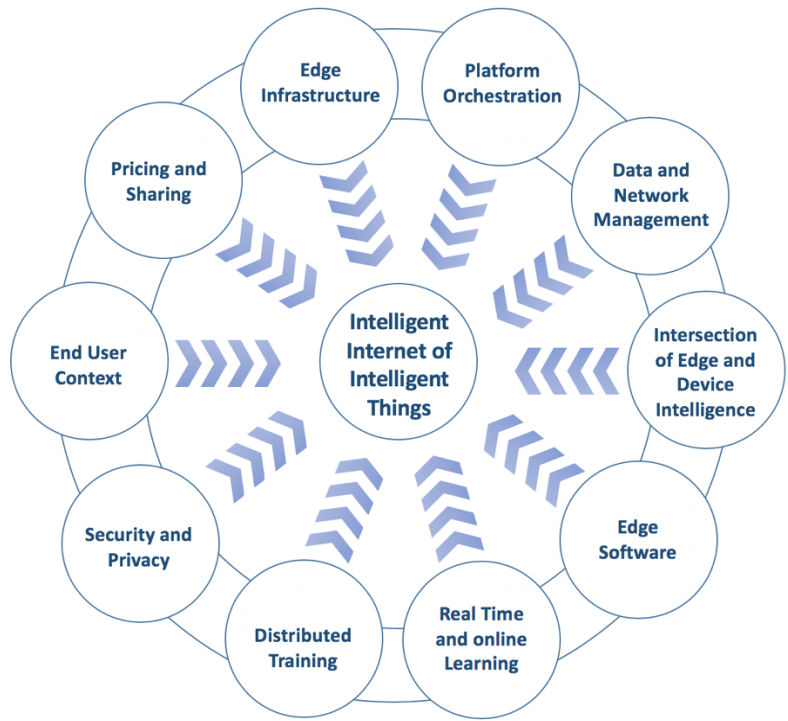

**FIGURE 3: KEY ENABLERS FOR INTELLIGENT INTERNET OF INTELLIGENT THINGS**

What we can easily anticipate is the fact that larger amounts of data will transit on the future 6G wireless communication network nodes and more and more added value applications and services will critically depend on that data. Bringing intelligence to the edge will clearly represent a basic functionality to guarantee the efficiency of future wireless communication networks in 6G and, at same time, can represent the enabling technology for a number of added value applications and services. **Artificial intelligence on the wireless communication nodes can actually enable a number of advanced services and quality of service functionalities for the proposed applications.**

Existing computing techniques used in the cloud are not fully applicable to edge computing directly due to the diversity of computing sources and the distribution of data sources. Considering that even those solutions available to transform heterogeneous clouds into a homogeneous platform are not presently performing very well, Mohan [6] investigates the challenges for integrating edge computing ((i) constrained hardware, (ii) constrained environment, (iii) availability and reliability, (iv) energy limitations) and proposes several solutions necessary for the adoption of edge computing in the current cloud-dominant environment. **We define that indeed, performance, cost, security, efficiency, and reliability are key features and measurable indicators of any AI for Edge and AI on Edge solutions.**

Zhou et al. [3] categorize edge intelligence into six levels, based on the amount and path length of data offloading. We extend Zhou et al.'s vision on edge-based DNNs to generic AI models and architectures and with seven levels where the edge can either be viewed as a set of single, autonomous, intelligent nodes or as a cluster or a collection of federated/integrated edge nodes. We also add a different degree of autonomy in the operation of the edge nodes (see Fig. 4). Specifically, our definition of the levels of edge intelligence is as follows:

- **Cloud Intelligence**: training and inferencing the AI model fully in the cloud.
- **Level-1: Cloud-Edge Co-Inference and Cloud Training**: training the AI model in the cloud but inferencing the AI model in an edge-cloud cooperation manner. Here edge-cloud cooperation means that data is partially offloaded to the cloud.

- **Level-2: In-Edge Co-Inference and Cloud Training**: training the AI model in the cloud but inferencing the AI model in an in-edge manner. Here in-edge means that the model inference is carried out within the network edge, which can be realized by fully or partially offloading the data to the edge nodes or nearby devices in an independent or in a coordinated manner.
- **Level-3: On-Device Inference and Cloud Training**: training the AI model in the cloud but inferencing the AI model in a fully local on-device manner. Here on-device means that no data would be offloaded/uploaded.
- **Level-4: Cloud-Edge Co-Training & Inference**: training and inferencing the AI model both in the edge-cloud cooperation manner.
- **Level-5: All In-Edge**: training and inferencing the AI model both in the in-edge manner.
- **Level-6: Edge-Device Co-Training & Inference**: training and inferencing the AI model both in the edge-device cooperation manner.
- **Level-7: All On-Device**: training and inferencing the AI model both in the on-device manner.

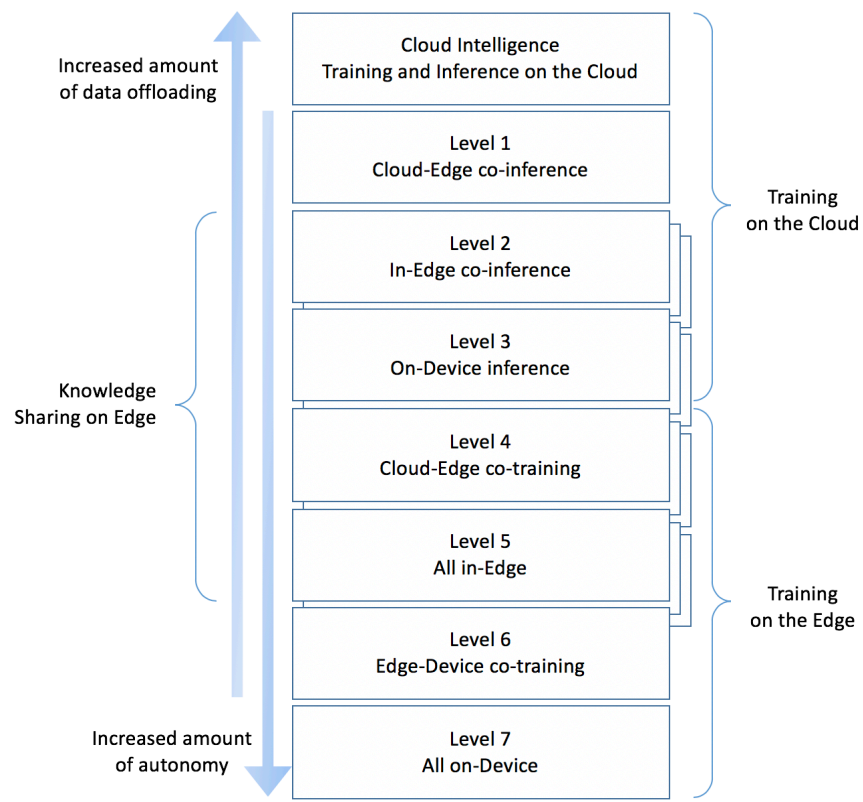

FIGURE 4: LEVEL RATING FOR EDGE INTELLIGENCE (ADAPTED FROM ZHOU ET AL [3])

Both AI for Edge and AI on Edge can be distributed at edge level. In practice, an edge node appears as a "local cloud" for the connected devices, and a "cluster of edge nodes" can cooperate to share the knowledge of the specific context and of the specific environment as well as to share computational and communication load both during training and during inferencing.

Further, we list and summarize a number of key functions that we envisage as useful for possible future edge intelligence applications at all possible levels of Fig 4. Therein, we highlight where exactly the Intelligence is "concentrated" and the applications and the services "are executed" depending on the specific application scenarios, on the local environment, the network architecture, the cooperative framework that can be defined, and the performance and the costs that need to be balanced. Some examples of Artificial intelligence methods to optimize telecom infrastructure in the 6G era and manage the life-cycle of edge networks (AI for Edge) are recalled in Table 1 Edge as a platform for application oriented distributed AI services (AI on Edge) are listed in Table 2.

It is worth nothing that to guarantee efficient, safe, secure, robust and resilient 6G based services it is also important to reduce dependencies between AI for Edge and AI on Edge services. Infrastructure and platform orchestration functionalities should guarantee their coexistence, and their optimization in case they coexist, but do not necessarily require those services to be fully implemented at all time. To allow for the maximum flexibility we should likely develop an "ontology for 6G connectivity" to shape all the possible combinations of "micro services" on the edge nodes.

| Table 1: AI for Edge Service | Specific Objective |
|---|---|
| Wireless networking | Zhu et al. (2018) [11] describe a new set of design principles for wireless communication on edge with machine learning technologies and models embedded, which are collectively named as Learning-driven Communication. It can be achieved across the whole process of data acquisition, which are in turn multiple access, radio resource management and signal encoding. |
| mmWave xhaul systems | Development of mmWave xhaul systems including AI/ML based optimization, fault/anomaly detection and resource management. Small Cells, Cloud-Radio Access Networks (C-RAN), Software Defined Networks (SDN) and Network Function Virtualization (NVF) are key enablers to address the demand for broadband connectivity with low cost and flexible implementations. Small Cells, in conjunction with C-RAN, SDN, NVF pose very stringent requirements on the transport network. Here flexible wireless solutions are required for dynamic backhaul and fronthaul architectures alongside very high capacity optical interconnects, and AI to maximize the collaboration between Cloud and Edge can represent a key solution. |
| Communication Service Implementation | Edge intelligence can automate and simplify development, optimization and run-time determination of communication service implementation. Edge intelligence in this case enables/assists service execution by determining the optimal/possible execution of service based on the resource availability in a network. |
| Dynamic task allocation | Offloading and onloading computational tasks and data between participating devices, edge nodes, and cloud, in addition to smart and dynamic (re-) allocation of tasks, could become the hottest topic when it comes to AI for edge. Dynamic task allocation studies the transfer of resource-intensive computational tasks from resource-limited mobile devices between edge and cloud, and interoperability of local devices sharing their computational power. These processes involve allocation of various different resources, including CPU cycles, sensing capabilities, available data and AI models, and channel bandwidth. Therefore, AI technologies with strong optimization and communication abilities can be extensively used in the 6G era. |
| Liquid computing handover | Seamless handover can be further extended to cover the handover of the tasks being shared between devices and Edge nodes while devices move in the network. |
| *Location-based Optimization* | Optimization of network coverage and of wireless networking can benefit a lot from the information collected progressively on the local radio environment. Basically, in this case, the devices exploit the knowledge available on the environment and on the Edge nodes. |
| *Predictive Quality of Service* | Quality of Service can be extended by predicting the behavior of the devices interacting with a specific Edge node or with group of Edge nodes. Information on the behavior and on the QoS can be shared between Edge nodes (with due attention to possible privacy and security issues). Basically, in this case, the Edge nodes exploit the knowledge available on the environment and the devices in that environment. |
| *Energy Management* | Although energy management for mobile devices has experienced significant improvements from the hardware point of view, we observe the high variability of energy performance of devices in connection with the applications. With the increasing level of autonomy that we expect for devices and for the 6G communication networks, we need to further extend the energy efficiency and energy management capacity for both 6G devices and 6G Edge nodes. |

| **Table 2: AI on Edge Service** | **Specific Objective** |
|---|---|
| Novel application areas | Autonomous and driving-assisted vehicles, autonomous drones, traffic control, smart factories, smart farms, smart roads, smart homes, smart cities, can actually define the reference profiles for services and for wireless communication network functionalities to be activated. |
| Data intelligence | Edge Intelligence is to use advanced communications technology and AI to support ubiquitous data collection, aggregation, fusion, processing, distribution and services at the Edge. The ability to learn, infer and control from data, in both static and dynamic environments, is an additional value-added feature. |
| Cooperative intelligence | Algorithms run on heterogeneous platforms which may be geographically distant (imposing latency requirements) jointly solving an AI problem. |
| Real-time requirements | Localized AI/ML functions with constrained computation resources and (usually) strict real-time requirements. |
| Computing as a Service | Provide intelligent computing capabilities when and wherever the user needs them (satisfying his/her requirements in terms of computing power, latency, energy consumption, cost, mobility, service reliability, etc.). |
| Advanced IoT models | IoT data models, architectures and smart services, especially with distributed services in IoT over different platforms and implementations. Here the focus is to enable services that can adapt based on available IoT devices/services in the network. Connected "smart objects" operate in an intelligent virtualized computational environment that is deployed across cloud, edge and mixed layers vertically and horizontally |

## 4. Challenges and key enablers

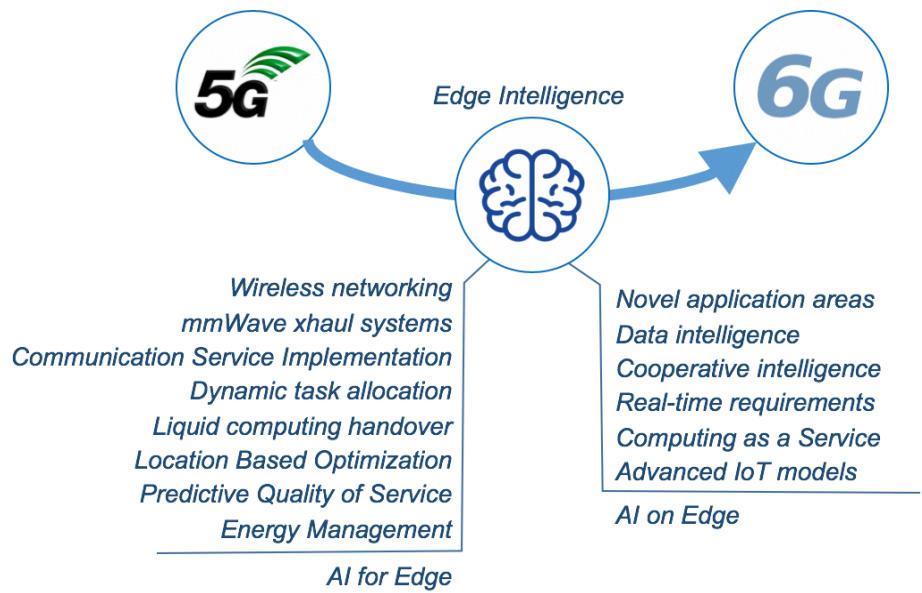

FIGURE 5: KEY CHALLENGES AND KEY ENABLERS FOR EDGE INTELLIGENCE FOR 6G

Although the benefits of edge intelligence are immense, the realization of the intelligence (training) in addition to the focus of applications (inference) poses several technical challenges in contrast to the traditional centralized artificial intelligence systems. Therefore, it is crucial to identify and analyze these challenges in the edge intelligence and seek for the novel theoretical and technical enablers. In this view, next, a set of prominent challenges in edge intelligence along some key enablers to overcome them are discussed.

### Edge infrastructure solutions

Edge computing infrastructures are best exemplified by the MEC reference architecture, currently under standardization by ETSI. The architecture describes edge platform components, their roles and expected functionalities, system APIs and interactions for collaboration and for 3rd party software integration. The target is an open multi-vendor edge platform, thus guidelines on how to realize systems and applications based on the reference architecture, and a set of proof-of-concept applications, are presented. Although the implementation details of the system components and interactions are left open, the architecture is based on distributed operation and control on two levels: system-level management and host-level management. However, the centralized orchestrator component is expected to have sole authority over all system resources. Platform- and host-level components operate based on instructions received from the orchestrator with partial autonomy to control resources under their domain. These components are expected to provide feedback to orchestrators about their operation. These operational principles are beneficial without a question but lead to challenges in real-time reactivity, providing low latency for multi-tenant applications, data routing and aggregation, system information delivery etc. across the dynamic and opportunistic distributed IoT environments. **Regarding AI capabilities for edge infrastructures, processing the system-wide data of resource usage and sharing across the deployment, system performance in relation to Key Performance Indicators (KPI), application data delivery, QoS and Quality of Experience (QoE) parameters etc. used for building models, learning, and further making predictions for optimization of the system behavior is largely unexplored territory.**

In architectural point-of-view, different approaches can be found. ETSI MEC is a two-tier architecture [12], with management components in the cloud, or platforms with similar capabilities, and application components deployed into the network edge layer below. Fog computing, in turn provides hierarchical computing platform across the deployment [13], typically confined in a space, where the computational units, i.e. Fog nodes have increasing capacity towards the cloud. Cloudlets are a similar concept, where on-demand physical computational capacity can be deployed in location as server racks and under/overused capacity shared across the platform, with additional cost of moving application components [14]. In such a multi-tier environment, the role of autonomous management and operation of local components is even more important and challenging. Recently, mist computing has emerged, where the data producing IoT devices, such as WSN nodes and mobile devices, are harnessed for application-specific data processing already at the data source. Another approach for device-level computing is mobile (cloud) computing, where the UE's of users provide ad-hoc shared computational capabilities at a location with offloading to the edge/cloud. In addition to MEC standardization, open-source solutions for edge platform management exist, such as Google Kubernetes and Docker Swarm that are widely used in industry.

### Edge platform orchestration

**The opportunistic nature of the IoT environment and large physical scale of edge computing systems justifies AI approaches for orchestration and management of such systems.** Further, optimization towards fulfilling the edge promises, e.g. efficient resource use and QoE, requires a large set of different data sources and complex data analysis algorithms. Such centralized algorithms would be initially difficult to design and develop, and later deploy, maintain and evaluate. Moreover, distributed and partitioned edge application execution is well aligned with the underlying architecture. A common challenge is to address the resource allocation problem, i.e., where to physically deploy edge computing infrastructure and what capabilities are needed in each location and its supporting logical "neighborhood" atop the physical network topology. A well-known approach here is to deploy the component next to existing infrastructure, e.g. wireless access points following the existing underlying network topology, or harness location-based low-resource computational units, as in Fog computing. Such deployments are limited by budgets and edge device capabilities, thus need to be carefully pre-planned, e.g. based on historical data, but also in response to online and predicted application workloads.

### Data and network management

**Availability, accessibility as well as the types of the data play pivotal roles in edge intelligence.** In contrast to the conventional centralized artificial intelligence, under most circumstances the concept of edge intelligence relies on "small data". Hence, generalizing the edge intelligence reliably over unseen data is a critical challenge. On the other hand, even when edge devices have a considerably large fraction of data, it is crucial to identify duplicates and anomalies to refine rich data to avoid performance losses (e.g. due to overfitting) in artificial intelligence models. Furthermore, the applications rely on edge intelligence may generate different types of data with multi-sensory (audio, video, haptic), spatial, temporal, and stochastic characteristics. Additionally, due to the fact that these heterogeneous sensory data is aggregated over a large network, the data itself may have inconsistencies. Thus, the fusion of these heterogeneous data types affects the edge intelligence performance.

In addition to data, the network states and requirements may change over time, even with extremely short durations (mission-critical applications) demanding tight response times. Under such changes, the trained artificial intelligence models need to have the capability of adapting or coping mechanisms. Sharing fractions of data and trained AI models instead of raw data will significantly reduce the communication payload size over any network, increasing the potential size of the data systems that can cope in a comparably short period of time. **Thus, it is crucial to understand and well-define data and model provenance and lifecycle and give measures to compare models and their fit to a current context on hand.**

Preprocessing data for machine learning is beneficial to govern efficient and reliable artificial intelligence models in the edge, cloud, and remote centralized data centers. Edge devices with large volumes of data can use clustering techniques to identify similarities therein and use tools of anomaly detection methods to isolate data inconsistencies. After the above classifications, down-sampling techniques can be adopted to manage the ensuring generalized edge intelligence. For the issues due to small data, edge devices need to increase sampling frequency with a cost of energy consumption to obtain rich data sets. Additionally, synthetic data can be generated by resorting to well-trained generalized adversarial networks. Furthermore, enabling incremental learning methods to train high quality models over time as well as adopting formally trained models via knowledge distillation and transfer learning techniques are promising solutions to cope with the issues posed by small data availability. Edge intelligence that needs to cope with the fusion of heterogeneous data types can utilize feature extraction techniques (i.e., representation learning) and split learning over multiple modalities. **To address the inconsistencies in heterogeneous sensory data, edge intelligence can resort to generalized adversarial networks, in which synthetic data could recover and restore the data consistency.**

Edge AI can be used to optimize the operations and performance of edge networks. However, in certain cases like fault detection and recovery, the problem is that failures are usually quite rare and thus data for machine learning is very imbalanced. In general, network traffic and events tend to have self-similar behavior, which means that, e.g., traffic anomaly detection mechanisms should be able to cope with heavy-tailed distributions. If user devices and applications start to adopt greedy AI based flow control and path/GW selection mechanisms, the network traffic flow might get even more harder to predict While several techniques can be used to generate artificial data, it is not necessarily clear that such data is optimal for training, e.g., fault detection systems. One alternative is to use simulations to obtain sufficient amounts of balanced data, which requires a kind of digital twin of the edge network. Furthermore, as the used RF frequencies are moving towards higher mmW bands (and over), it might be beneficial to integrate also other types of data to network management, like fine-grained localized weather information (heavy rain events) and seasonal changes in foliage. **To summarize, it would be better to understand the behavior of the network instead of handling it as a black box.**

### INTERSECTION OF EDGE AND DEVICE INTELLIGENCE

**The location of the edge application has utmost importance regarding real-time reactivity and adaptivity in response to the dynamic environment and user movement.** To address the extreme end of distributed edge, edge-supported approaches including mobile devices as a part of the computational platform have since emerged, including Mobile Cloud Computing, Mobile Edge and Mobile Fog. Naturally, such platforms at the low end possess limited capabilities for "small data" processing, analysis and dissemination, where further support from edge is required for advanced analysis. Therefore, such distributed applications are typically partitioned with software components on the user devices, and edge and cloud layers. Here, sharing of (refined) data and local resources both horizontally and vertically has utmost importance to save device resources, i.e. energy, on the participating devices and to provide operational capacity in response to user mobility. Key challenge is interoperability, i.e. uniform interfaces, to share data, results, tasks and high-level AI models (e.g. algorithms). **Moreover, the massive scale of such distributed deployments across networks significantly increases the scale of management and orchestration with a holistic view over system operation and introduces further latencies into control.** Therefore, lightweight AI solutions are needed already at the mobile device level to increase autonomy and self-* capabilities.

Here a classical distributed AI paradigm, software agents and multi-agent systems, have shown benefits in providing autonomy, reactivity, adaptivity, machine learning, code mobility and collaboration capabilities [15, 16], and even for the resource-constrained IoT devices [17]. Such devices are commonly known as smart objects in the IoT context [18, 19]. Existing use cases for agents in cloud-edge-device continuum include representing system entities, facilitating collaboration both horizontally and vertically, and sharing of resources, and controlling (e.g. SDN) and monitoring the system operation, networks and devices. **However, how to increase the agent capabilities from reactive operation towards deliberative agents with cognitive capabilities, e.g. learning and proactivity, is an open question.** Further AI techniques, facilitating both vertical and horizontal collaboration and cooperation include swarm intelligence, game theory and genetic algorithms.

### Software development for edge

**Software development for edge systems relies on virtualization, exemplified by virtual machines and lightweight containers.** Edge applications are developed as software packages, possibly implemented by multiple stakeholders, from where the application images are automatically built by edge system management components that maintain the application lifecycle. The images are then deployed to the edge hosts atop the virtualization infrastructure according to the system policies and further managed and instantiated by platform and host components, which are additionally responsible for providing the required application-specific service, data and network access and maintaining the required QoE. The challenge, in addition to deployment considerations, here is to manage system policies, SLAs, access rights, billing, etc, for the software packages and negotiate and orchestrate their use online, possible with external 3rd party service providers and network operators. In deploying and launching edge applications, both push and pull approaches are facilitated, through offloading from UE and pulling application-specific components to near infrastructure. **Also, towards autonomy, horizontal code migration of selected software components, in a limited scope, is an enabler.** In this context, security and resource and information sharing are important challenges.

Virtual machines typically provide monolithic self-contained application images, typically in size of several gigabytes, that become resource-consuming to deploy and move across the edge platform. A distributed approach for edge application development is microservices, where the goal is to develop modular application components, in individual process level, in isolation that can be individually deployed on-demand to build the application workflow. **Here, a lightweight version of virtual machines, containers, encapsulates individual microservices for deployment.** Unikernels that can run as virtual machines or even at bare metal, provide an alternative to containers. The image size of unikernel based microservices can be more or less the same size as container-based alternatives and thus unikernels could provide better isolation with the same resources.

DevOps practices, such as CI/CD, provide means to develop microservices in isolation, managing versioning and deploying such components automatically by system management components. Obviously, managing such large-scale automatization on-line is challenging, since the additional small-scale application-specific units and their workflows increase significantly the scope of system and package management and related performance monitoring. There are some AI specific CI/CD frameworks, like Kubeflow[9] and MLFlow[10], that support AI model development, training and deployment workflows in cloud environments. However, in edge AI environments, such frameworks would need adaptation to edge data sources and federated cloud environments.

### Real-time requirements and online learning

Novel and future AI applications require real-time feedback to be effective and address challenges set by many real-world applications, including robotics and self-driving cars, traffic and logistics management systems, and telepresence, virtual and augmented reality applications all included to the 6G verticals and application areas. Real-time challenges cannot be solved only by decreasing latency and increasing network bandwidth due to time usually spent to collect the data for machine learning models, training these models, and defining actions based on the learned models to be returned to the application. **Thus, re-defining the whole real-time feedback cycle becomes even more crucial including balance between pre-trained and online learned models, efficient model distribution and re-utilization during their lifecycle, and dynamic decision making based on all the knowledge available from different models and data sources.**

---

[9] https://www.kubeflow.org/docs/started/kubeflow-overview/
[10] https://mlflow.org/

To address the challenges due to the need of short response time, it is essential to quickly adapt data dissemination and model training along the network changes as well as to reduce the processing complexity in the inference. By using the frameworks of transfer learning and knowledge distillation, edge intelligence can reduce retraining latency with the aid of pre-acquired intelligence. Furthermore, knowledge distillation and model pruning allow to reduce the artificial intelligence models yielding fast inference. In addition to aforementioned methods, the dynamics in the data and the network can be addressed by resorting to reinforcement learning and the codesign of communication, control, and machine learning [20].

### DEVELOPING DISTRIBUTEDLY TRAINED ALGORITHMS

Towards realizing edge intelligence, the training procedure directly affects the majority of the end-to-end latencies, the inference reliability, and the overall scalability [8]. While a handful of applications may allow traditional centralized artificial intelligence model training and download trained-model for the inference at the edge, **the majority of mission-critical and privacy-concerned applications demands for online distributed training algorithms that can be employed at the edge devices**. In this view, on-device limitations and the communication bottleneck among edge devices as well as between the edge and the servers play a critical role in developing the distributed algorithms. The edge devices in a large-scale artificial intelligent system are likely to be mobile and thus, powered with capacity-limited batteries and storage. The limited energy budget is used for both the computation (training and inference) and the communication.

While large machine learning models and frequent coordination among edge devices are preferred for higher inference accuracy and reliability [21], they could be inefficient from the energy consumption prospects, bounded by the storage/memory limitations as well as the privacy requirements. These on-device constraints call for energy-efficient, low-complexity and capacity, privacy-sensitive designs of distributed algorithms. Under the limited data availability, edge devices may require exchanging raw data itself, their model parameters, or inferred outputs/decisions among one another or with a central server to improve the reliability and the robustness of the distributed algorithms. This coordination within the network is suffered by the uncertainties in the communication links and the network dynamics. **Therefore, user and resource (computation and communication) scheduling as well as data and model compressing need to be accounted for in the distributed algorithm design.**

The aforementioned limitations of device capabilities and communication in distributed algorithm design can be addressed with several technical and theoretical enablers listed below. Under the limited power and memory on devices, it is suitable to seek for data and model parallelization techniques depending on the privacy requirements. Here, data can be split into several batches and processed utilizing the mobile edge computing servers. Alternatively, large artificial intelligence models can be split over several devices in which sequential/parallel training can be carried out, i.e., adapting split learning within distributed training algorithms [22]. Additionally, depending on the training data size and privacy requirements as well as the artificial intelligence model size, federated learning, knowledge distillation, and transfer learning methods can be selected as model training techniques [23, 24, 25].

Towards improving the communication efficiency within distributed training algorithms, the uplink-downlink channel capacity asymmetry in the wireless network can be exploited by jointly adapting knowledge distillation and federated learning. Moreover, artificial intelligence model pruning, coded and quantized model/data transmission-based learning can be adopted to address the limitations of both communication and storage [26]. **Furthermore, it is important to identify the characteristics of the network dynamics while designing the distributed training algorithms.** Since static algorithms may yield poor performance under the drastic changes in the network, it is crucial to either introduce cold-start mechanisms to retrain the models or to adopt continuous knowledge acquisition via continual learning methods including transfer learning, online learning, and reinforcement learning. In addition to aforementioned technical enablers, it is mandated to ensure latency, reliability, and scalability guarantees within the distributed training algorithms. In the view of reliability, generalization error is a performance measure of a trained artificial intelligence model over unseen data. For distributed algorithms, the frameworks of meta distribution, risk management, and extreme value theory can be used as theoretical enablers to analyze and minimize the generalization errors. To reduce the latency while developing secure and scalable distributed algorithms, tools from differential privacy, rate-distortion theory, and mean-field control theory can be adopted.

### Security and privacy

In any edge-cloud computing environment data is transmitted from the edge to the dedicated computing infrastructure with services that will perform the data analysis which can be either private or public. Since the data leaves the edge, it can be exposed to various vulnerabilities and attacks such as penetration attacks resulting in theft of information or even Denial of Service attacks resulting in crashing servers or networks. Additionally, an attacker can not only access and intercept the data, but also the application's processing outcome in transit to lead to a different action/scenario than the intended one (i.e. tampering). **Additionally, the locality on the edge, as well as the potential proximity of the system to end users can enable it to help address certain security challenges.** In some applications, edge AI can also be used to improve security and privacy, e.g., by anonymizing human faces in video streams or replacing people with straw figures if the main application depends only on the number of people in a certain location or if someone has fallen and is lying on the ground.

**In this view, lightweight and distributed security mechanism designs are critical to ensure user authentication and access control, model and data integrity, and mutual platform verification for edge intelligence.** Also, it is important to study novel secure routing schemes and trust network topologies for edge intelligence service delivery when considering the coexistence of trusted edge nodes with malicious ones. On the other hand, the end users and devices would generate a massive volume of data at the network edge, and these data can be privacy sensitive since they may contain user's location data, health or activities records, or manufacturing information, among many others. Subject to the privacy protection requirement, e.g., **EU's General Data Protection Regulation (GDPR),** directly sharing the original datasets among multiple edge nodes can have a high risk of privacy leakage. Thus, federated learning can be a feasible paradigm for privacy-friendly distributed data training such that the original datasets are kept in their generated devices/nodes and the edge AI model parameters are shared. To further enhance the data privacy, more and more research efforts are devoted to utilizing the tools of differential privacy, homomorphic encryption and secure multi-party computation for designing privacy-preserving AI model parameter sharing schemes.

### End-user aspects

One of the main goals of edge computing, and justification for edge intelligence, is to maintain required QoE for users, in terms of network connectivity and application execution improvements, and adaptation to the dynamic environment and user mobility. **A key challenge in optimizing the edge systems towards QoE is understanding the user context (e.g. location-awareness),** based on both large-scale analysis of user behavioral patterns across the deployment, but also real-time reactivity in the local environment (e.g. in relation to network connectivity and latencies, bandwidth availability, data transmission and application execution requirements and integration of user-specific third-party components). These concerns lead to online and on-demand adaptation of local edge resources, that propagates across the deployment both horizontally and vertically.

**Challenges are introduced by user mobility**, leading to user- and application-specific (virtualized) component movement and migration across the edge deployments, and to management challenges in data and (stateful) application migration while minimizing latencies of handovers. Further, the dynamic edge resources are shared between multiple users in multi-tenant fashion, raising privacy and security concerns. Regarding mist computing, sharing user equipment as a part of the data collection and computational platform requires incentives to encourage participation, such as micropayments, defining data ownership(s) and policies for sharing, and privacy protection schemes with GDPR compliance. Approaches for representing users in edge systems have been already proposed, e.g. digital twins and software agents with cognitive capabilities. Here, building trust between edge platforms and users is required for successful cooperation.

#### PRICING AND SHARING MECHANISM

In future 6G networks, the AI-powered mobile edge devices are enabled to share their communication, caching, computation, and learning resources (3C-L resources) to satisfy the quality of requirements (QoE and QoS) for 6G wireless applications, such as Tactile Internet, virtual reality, and autonomous driving. Hence, an intelligent 3C-L resources sharing framework is still at its infancy and should be significantly addressed, where all the resources can be shared by the mobile edge devices to maximize the resources utilization by virtualizing all the resources into the virtual resource pool.

To cope with such a challenge, the 3C-L resource sharing can be modeled by a dynamic pricing mechanism from an economic perspective, in which the mobile edge devices are modeled as intelligent agents that can price, i.e. operate as brokers, or purchase 3C-L resources and consume services according to their own will. Accordingly, an economic sharing model should be set up to make 3C-resources and knowledge tradable by market equilibrium approach. In particular, the multi-agent distributed learning approach can be developed to make the optimal price and resource allocation decisions considering different QoE and QoS requirements of 6G network applications, services and systems. Moreover, how to smartly record the price and disseminate the revenue according to the proof of work among the distributed edge devices is also very important. **Smart contracts and distributed ledgers are expected to take an important role in fully unleashing the fairness, security and activity of this ecosystem.** Therefore, designing the proper sharing and incentive mechanisms, as well as lightweight consensus protocols for edge intelligence should elicit escalating attention.

Table 3 summarizes the defined key challenges and enablers.

| Table 3: | Key challenges | Key enablers |
|---|---|---|
| **Edge computing infrastructure** | Support for dynamically changing resources and configurations<br>Reliable feature deployment<br>Device mobility | Container technologies<br>Virtualization<br>Isomorphic software architectures<br>Handover protocols and techniques |
| **Edge platform orchestration** | Address the resource allocation problem.<br>Address resource distribution problem.<br>Address dynamic allocation and distribution problems. | Virtualization optimization.<br>Data intelligence algorithms.<br>Data analysis algorithms.<br>Multilevel and fully distributed dynamics optimization algorithms. |
| **Data and network management** | Fusion of heterogeneous data types to optimize edge intelligence performance.<br>Understanding and definition of data and model provenance and lifecycle.<br>Compare models and their fit to a given context.<br>Enable incremental learning methods.<br>Address inconsistencies in heterogeneous sensory data.<br>Modelling and understanding network behavior and performance.<br>Common Interoperability practices and standards optimized high-level AI models.<br>Centralized system management and operation, limiting control latencies | Extension of sensor and data fusion algorithm to support proper data and network management.<br>Data and network fault detection and identification.<br>Knowledge extraction and incremental learning algorithms<br>Knowledge sharing solutions.<br>Lightweight AI solutions to increase autonomy and self-* capabilities.<br>Agents with cognitive capabilities, e.g. learning and proactivity.<br>AI techniques for collaboration and cooperation, adopt well-known paradigms such as swarm intelligence, game theory and genetic algorithms. |
| **Intersection of Edge and Device Intelligence** | Common Interoperability practices and standards and optimized high-level AI models.<br>Centralized system management and operation, limiting control latencies | Lightweight AI solutions to increase autonomy and self-* capabilities.<br>Agents with cognitive capabilities, e.g. learning and proactivity.<br>AI techniques for collaboration and cooperation, adopt well-known paradigms such as swarm intelligence, game theory and genetic algorithms. |
| **Software development for Edge** | Dynamic configurations<br>Security<br>Flexible deployment<br>Debugging capabilities in development time | Container technologies<br>DevOps, CI/CD<br>Virtualization<br>Liquid software that can flow from one node to another |
| **Real-time requirements and online learning** | To adapt along the network dynamics<br>The need of short response times | Transfer learning, knowledge distillation and reinforcement learning<br>Codesign of communication, control and machine learning |
| **Distributedly trained algorithms** | Reducing the cost of coordination among edge devices<br>Lowering generalized errors in trained models | Federated learning, knowledge distillation, transfer learning<br>Model pruning coded and quantized machine learning<br>Meta distributions, extreme value theory, risk management framework |
| **Security and privacy** | Guarantee the implementation of security and privacy strategies according to user needs and requirements.<br>Guarantee the recognition of abnormal behavior according to user requirements and to operator criteria. | Allow the application of the security strategy at physical level. Use physical level solutions to increase security.<br>Allow the proper management of personal information ownership at all stages. |
| **End-user aspects** | Understanding user context based on both large-scale behavior patterns as | Digital twins and software agents with cognitive capabilities |

# 5. Core Research Questions

The emergence of IoT and the demand for responsiveness, privacy, and context-awareness are pushing intelligence to the edge and are pulling those challenges and enablers listed in Table 3. For numerous domains that utilize and benefit from 6G, Edge Intelligence can be outlined with reference to the services to enable "AI for Edge" of Table 1 and to the services to enable "AI on Edge" of Table 2. Next, we provide an overview of the core research questions that shall be tackled in the development of 6G Edge Intelligence.

## Edge Infrastructure

1. What architectural considerations are involved in application development with edge intelligence? Flexible architecture and pipeline to cater for different types of edge devices? Seamless vertical and horizontal collaboration in physical deployments atop existing network topologies?
2. How to define edge dedicated classification and taxonomy for the edge intelligence components?
3. What are the impacts on network architecture regarding outer-network edge intelligence and in-network edge intelligence, including Network Service and Function (NSF) -level intelligence and inter-NSFs intelligence. How to develop suitable network protocols and interfaces for edge intelligence?

## System Platform and Stack

4. What software development practices, quality assurance and testing methods could be applied with edge intelligence-powered applications?
5. DevOps aspects?
6. How to enable diagnostics and cooperative diagnostics for edge specific algorithms? How to involve multiple edge entities to debug in a decentralized setting?
7. How to incorporate formal verifications tools and processes wherever possible?

## Data and Network Management

8. How to organize online learning with (possibly) non-stationary data, federated learning methods, energy-efficiency?
9. How to design agent-based systems to enable cognitive capabilities in real-world?
10. How to distribute computing among the different resources?

## Intersection of Edge and Device Intelligence

11. How to implement lightweight AI solutions to increase autonomy and self-capabilities?
12. How to include agents with cognitive capabilities?
13. How to optimize AI techniques, facilitating both vertical and horizontal collaboration and cooperation and including swarm intelligence, game theory and genetic algorithms?

## Software Development of Edge

14. How to develop flexible and interoperable software agents?
15. How to guarantee reconfigurability and continuous deployment?
16. What solutions can be adopted for maximum virtualization capacity?
17. How to guarantee liquid software solutions that can flow from one node to another?

## Real-time requirements and online learning

18. How to guarantee the capacity to transfer learning, and provide knowledge distillation and reinforcement learning?
19. How to implement codesign of communication, control and machine learning?
20. How to quickly adapt model training along the network changes?
21. How to reduce the processing complexity in the inference?

### Distributed Training, Algorithmic Design and Deployment

22. What is the impact of accelerators and how they are changing the way we approach distributed algorithms and ML design?
23. How to provide enough (and powerful) resources for ML at the edge (e.g. hardware acceleration, GPUs, etc.) in an economically sustainable way?
24. How to balance the energy required for performing the computation under different scenarios? Enabling Computational Power/Storage Capacity/Power Consumption on the edge?
25. How do carry out distributed training, inference and control in a communication-efficient, reliable and scalable manner?
26. How to factorize ML algorithms that can run partly on heterogeneous platforms (with heterogeneous computing infrastructural resources also? For example, training deep neural networks (estimation of parameters)?

### Security, Privacy, and Portability

27. Security of data, of ML model tempering, of protocols (new protocols must be invented that use TPMs at the edge devices)
28. How to make operation "fail safe" in case of network failure, node failure
29. How to ensure data provenance at scale?
30. How to ensure that the privacy concerns of users and regulatory bodies are satisfied?

### End-user Concerns

31. Address QoE related KPIs of edge standardization
32. How to share resources among the different end users?
33. How to handle user mobility?
34. How to handle stateful application migration?
35. What are the social-technological influences of edge intelligence
36. What are the regulation or cultural issues involved?
37. How to motivate the owner, the developer, the operator, the tenant, and the user of edge intelligence, then to define the relationship among them and the relationship with those of the network?

### Pricing and Sharing mechanism

38. How to virtualize the 3C-L resources?
39. How to set up an economic sharing model by market equilibrium approach?
40. How to design the dynamic pricing mechanism?
41. How to smartly trade 3C-L resources and disseminate the revenue?
42. How to design the proper sharing incentive mechanism?

# 6. Prospective Use Cases

Edge Intelligence methods will provide novel business opportunities and technological solutions for various application fields, including but not limited to, personal computing, urban computing, and manufacturing easing their efficient, safe, secure, robust, and resilient wireless networking. Next, we highlight some example application areas for Edge Intelligence and especially Intelligent Internet of Intelligent Things.

## Edge Intelligence for Autonomous Driving

Autonomous driving is one of the broad research topics covering applications ranging from simple driver assistance systems (such as e.g., traffic signs recognition, lane keeping assistance, etc.) to fully automated driving without human support. One of the specific and interesting use cases of autonomous driving is the concept of autonomous vehicle platooning, i.e., a coordinated movement of a group of driverless cars. This group of short-distance vehicles, typically trucks, forms a convoy, which is led by a platoon leader, responsible for sending the steering information to the platoon members.

The data exchange between the platoon members and between the platoon and other devices can be realized by means of different wireless communications technologies, (such as the Dedicated Short-Range Communications, DSRC, or cellular networks, Cellular-V2X, C-V2X), however when the number of communicating cars increases, it may suffer from the prospective medium congestion and might not be able to fulfill the stringent requirement of 99.99% reliability. In this context, the approach to offload some data to other bands is gaining interest. EI is foreseen here as the enabler for dynamic spectrum access that should allow for fast and reliable processing of data generated within platoons and by all users of the street. A hierarchical structure of the database-oriented system supporting these operations in V2X communications, as shown in Fig. 6. We consider a motorway scenario, where multiple platoons of cars travel among other vehicles. We assume that platoon cars are autonomous with their mobility controlled using the Cooperative Adaptive Cruise Control (CACC) algorithm.

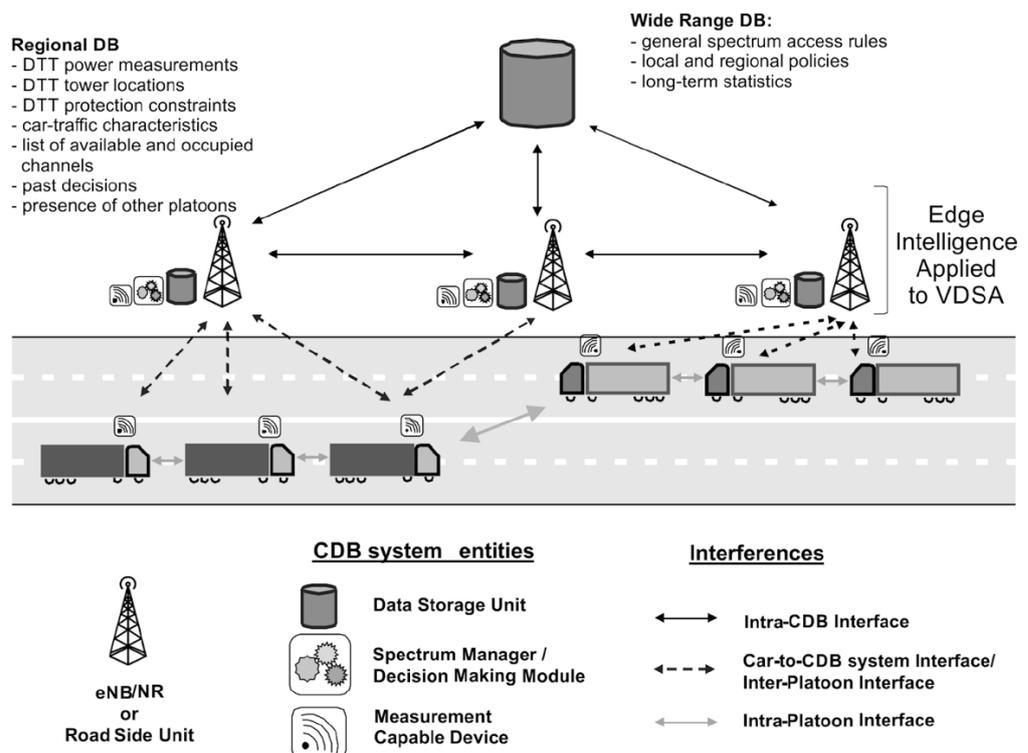

FIGURE 6: Edge Intelligence Applied to vehicular dynamic spectrum access in platoons

The intra-platoon communications in dynamically-allocated band as a secondary system should not cause the degradation of any existing licensed service. In order to support vehicular dynamic spectrum access, the database-oriented systems equipped with dedicated units for data processing and decision making shall specify which frequency bands can be used for data transmission. We claim that various kinds of information will be stored in local (regional) and global databases. Edge intelligence functionality will be provided by the dedicated advanced processing units, that are co-located with the base station or road side units.

### EDGE INTELLIGENCE FOR SMART SPACES

Smart spaces such as smart homes, smart campuses, smart offices, and smart hospitals are expected to contain a variety of networked devices and AI-driven in-network services to aid with everyday activities. These devices and services will be latency-sensitive, they are expected to exchange privacy sensitive information, and some of these devices, such as surveillance cameras, are expected to generate large volumes of data. Satisfying the requirements of applications that use these devices and services will require edge-native solutions. For example, data generated by the networked devices in smart hospitals can be privacy sensitive. Regulations such as GDPR may mandate storing and processing of the data within the hospital premises. Similarly, AI services for object and face recognition from live surveillance footage in smart spaces require real time processing of large volumes of data, motivating the need for edge-native solutions.

### EDGE INTELLIGENCE FOR ENVIRONMENTAL SENSING

Environmental sensing such as air quality monitoring requires collecting and processing large volumes of data from a variety of sensors spread across large geographical areas [27]. For instance, accurate air quality monitoring demands a high spatial and temporal resolution of air quality data from sensors that monitor humidity, temperature, particulate matter concentrations, and gaseous pollutants. Collecting this data requires a dense deployment of sensors and real-time processing and filtering of the raw analog data collected by the sensors. AI can be leveraged for identifying the optimal locations for sensor deployment, the trajectories for the mobile sensors, the calibration of the low-cost sensors, and the locations for performing the computation. Edge-native solutions for processing the raw data can help reduce the network load and also parallelize the computation.

### EDGE INTELLIGENCE FOR MOBILE XR

Mobile extended reality (XR), a portfolio encompasses all virtual or combined real-virtual environment compounds including virtual reality (VR), augmented reality (AR) and mixed reality (MR), is a promising AI-powered service application of 6G (e.g., future Tactile Internet). Visuo-haptic XR allows remote interaction with real and virtual elements (objects or systems) in perceived real time and drives massive real-time data at the network edge. Therefore, it is a kind of computation-intensive and data-craving application with low latency supporting. Edge intelligence has demonstrated great potential especially for XR service on resource and energy-constrained devices. To address the limitations of battery energy of devices, computation capacity and reduce end-to-end latency, Intelligent task segmentation, computing offloading and learning model sharing will play a great role in bringing immersive experiences for users.

### NEXT GENERATION COBOTS (COLLABORATIVE ROBOTS) IN MANUFACTURING

The collaboration between robots and humans in various domains will increase and become more seamless in the future. In manufacturing, not only are routine tasks being transferred from humans to robots (while upskilling people), but collaborative robots also perform task-level collaboration with humans. In their current state, cobots are stationary devices with fixed gripper mechanisms and task programming. In the future, cobots are envisioned to have the following functionalities: Automatic monitoring of machine health properties, autonomous or semi-autonomous navigation within the factory floor, switching from workstation to another through task-level adaptation, and collaborating as a fleet. These functionalities call for various sources of real-time data generation to cobots themselves, as well as low latency communications and tight collaboration with MES (Manufacturing Execution System) systems and factory private clouds. This again calls for a role for edge intelligence as performing fine-grained control of cobots, as well as coordinating larger production goals with back-end MES/cloud.

## 7. Roadmap to Edge Intelligence

Edge Computing is one of the key technologies that is making it possible for 5G networks to satisfy the stringent requirements in different use cases such as URLLC (contributing to low latency) or mMTC (providing distributed computing power). AI goes even beyond, becoming essential techniques in the technology industry for implementing a wide variety of applications such as video processing, data analysis, image generation, and so on. Thus, as shown in the "Related work" section of this white paper, there is no doubt that both technologies will be combined into Edge Intelligence to play an important role in 6G.

**The evolution to the deployment of a new generation of edge intelligence systems, applications and services will take place during the next ten years, with the completion of different technological steps that will provide new devices, technology and applications, as shown in the roadmap below.**

There are several challenges that have to be addressed. For example, hardware needs to evolve to make it economically viable the deployment (at the edge) of a large number of efficient architectures and devices supporting existing and new AI techniques with high computing requirements, such as DNN, and supporting also the larger amounts of data that will transit on the future 6G wireless communication networks. Software will also advance in different aspects such as distribution, automation, intelligent orchestration of components, security, etc.

Thus, we expect a steady, sustained work that will address the main challenges identified in this white paper, providing new results in the different technological aspects related to edge intelligence. For example, the first deployments will use pre-trained models, but this will have to progress into systems that combine pre-trained and online learned models that will be able to define actions based on the information collected, even in real-time. Despite the increasing power of edge hardware, it will be necessary to distribute both training and data processing. New hardware components (e.g. new AI accelerator application-specific integrated circuits) will be developed to improve the performance while reducing the energy consumed and the cost. In the long term, technologies such as nanophotonics may be used for performing complex operations or store information. Communications and learning will be combined and exploited by learning-driven design principles. In addition, by taking advantage of the distributed nature of the edge architecture, new proposals will arise to keep the data and intellectual property secure, and to ensure user privacy.

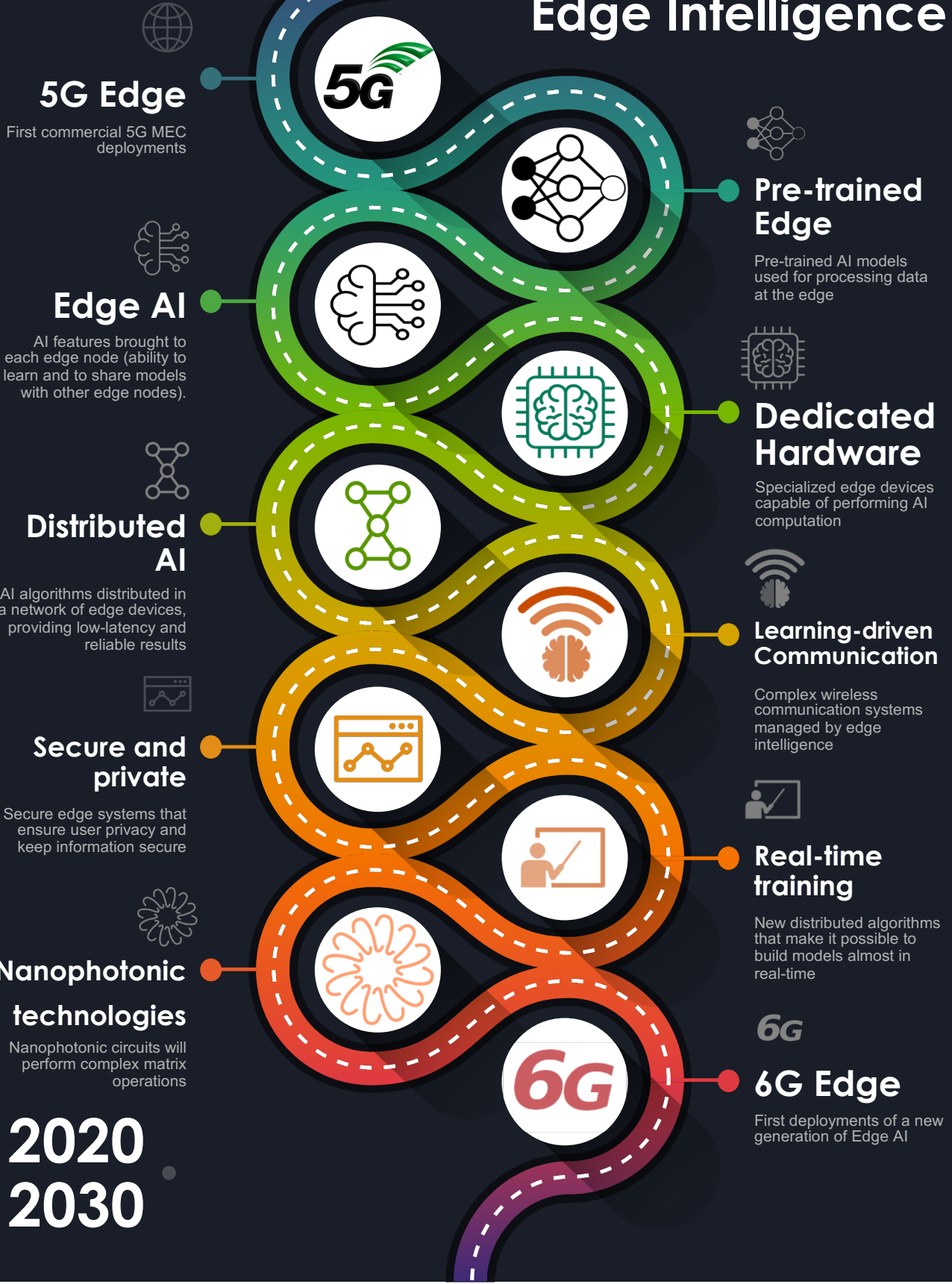